\documentstyle[11pt]{article}
\newcommand{\be}{\begin{equation}}
\newcommand{\ee}{\end{equation}}

\newcommand{\ts}{\hspace{3pt}}

\begin{document}

\vspace*{1.5cm}

\noindent
{\LARGE \bf  	The Wave Function: It or Bit?\footnote{\ Draft of
an invited chapter for {\it Science and Ultimate Reality}, a
forthcoming book in honor of John Archibald Wheeler on the occasion of
his 90$^{th}$ birthday (www.templeton.org/ultimate\_ reality).} }

\vskip 1.5cm
\begin{quote}
\noindent
{\bf H. D. Zeh}
\vskip 0.2cm
\noindent
Universit\"at Heidelberg\\
www.zeh-hd.de
\end{quote}
\vskip 1.8cm

{\bf Abstract:} Schr\"odinger's wave function shows many aspects of a
state of incomplete knowledge or {\it information} (``bit''): (1) it
is usually defined on a space of classical configurations, (2) its
generic entanglement is, therefore, analogous to statistical
correlations, and (3) it determines probabilities of measurement
outcomes. Nonetheless, quantum superpositions (such as represented by
a wave function) define individual {\it physical} states
(``it''). This conceptual dilemma may have its origin in the
conventional {\it operational} foundation of physical concepts,
successful in classical physics, but inappropriate in quantum theory
because of the existence of mutually exclusive operations (used for
the definition of concepts). In contrast, a {\it hypothetical
realism}, based on concepts that are justified only by their universal
and consistent applicability, favors the wave function as a
description of (thus nonlocal) physical reality. The (conceptually
local) classical world then appears as an illusion, facilitated by the
phenomenon of decoherence, which is consistently {\it explained} by
the very entanglement that must dynamically arise in a universal wave
function.

\vfill

\section{Introduction}
Does Schr\"odinger's wave function describe physical reality (``it''
in John Wheeler's terminology \cite{timeas}) or some kind of
information (``bit'')? The answer to this question must crucially
depend on the definition of these terms. Is it therefore merely a
matter of words? Not quite -- I feel. Inappropriate words may be
misleading, or they may even be misused in the place of lacking
explanations, while reasonably chosen terms are helpful.

A {\it bit} is usually understood as the binary unit of information,
which can be {\it physically realized} in (classical) computers, but
also by neuronal states of having fired or not. This traditional
physical (in particular, thermodynamical) realization of information
(``bit from it'') has proved essential in order to avoid
paradoxes otherwise arising from situations related to Maxwell's
demon. On the other hand, the concept of a bit has a typical quantum
aspect: the very word quantum refers to discreteness, while,
paradoxically, the {\it quantum bit} is represented by a continuum
(the unit sphere in a two-dimensional Hilbert space) -- more similar
to an analog computer. If this quantum state
describes ``mere information'', how can there be {\it real} quantum
computers that are based on such superpositions of classical bits?

The problematic choice of words characterizing the nature of the wave
function (or ``quantum state'', in general) seems to reflect the common
uneasiness of physicists, including the founders of quantum theory,
about its fundamental meaning. However, it may also express a certain
prejudice. So let me first recall some historical developments,
most of which are described in Max Jammer's informative
books \cite{Jammer}, where you will also find the relevant ``classic''
references that I have here omitted.

\section{Historical Remarks about the  Wave Function}
When Schr\"odinger first invented the wave function, he was
convinced that it  describes real electrons, even though the
construction of his wave equation from Hamiltonian mechanics readily
led to wave mechanics on {\it configuration} space. As far as he
applied his theory to single electron states, this had no immediate
consequences. Therefore, he tried to explain the apparently observed
``corpuscles'' in terms of wave packets in space (such as coherent
oscillator states). This attempt failed, but I will apply it to the
configuration space of bounded systems in Sect.\ts 4. Since
Schr\"odinger firmly believed that reality must be described in space
and time, he proposed nonlinear corrections to his single-electron
wave equation, thus temporarily abandoning his own many-particle wave
function.

When Born later proposed his probability
interpretation, he initially postulated probabilities for
spontaneous transitions {\it of a wave function into a new one}, since
at this time he ``was inclined to regard it [wave mechanics] as the
most profound formalism of the quantum laws'' (as he later explained).
These new wave functions were either assumed  to be plane waves (in a
scattering process), or bound states (for
spontaneous transitions within atoms). In both cases the
final (and mostly also the initial) states were stationary eigenstates
of certain subsystem Hamiltonians, which thus replaced Bohr's
semi-quantized electron orbits of the hydrogen atom.\footnote{\ The
idea of probabilistically changing (collapsing) wave functions was
generalized and formalized as applying to measurements by von Neumann
in what Wigner later called the ``orthodox interpretation'' of quantum
mechanics. (By this term he did {\it not} mean the Copenhagen
interpretation.) Its historical roots may explain why
von Neumann regarded quantum jumps as the {\it first} kind of
dynamics, while calling the Schr\"odinger equation a {\it
second} ``Eingriff'' (intervention).}   Born ``associated'' plane waves
with particle momenta according to de Broglie's mysterious proposal,
although this had already been incorporated into wave mechanics in the
form of differential momentum operators. Only after Heisenberg had
formulated his uncertainty relations did Pauli introduce the {\it
general} interpretation of the wave function as a ``probability
amplitude'' for particle positions {\it or} momenta (or functions
thereof) -- cf.\ts
\cite{Beller}. It thus seemed to resemble a statistical
distribution representing incomplete knowledge (although not about
position {\it and} momentum). This
would allow the {\it entanglement} contained in a many-particle
wave function to be understood as statistical correlations, and the
reduction of the wave function as a ``normal increase of
information''.

However, Pauli concluded (correctly, I think, although this has also
been debated) that the potentially observed classical properties
(particle position or momentum) cannot be merely unknown. Instead, he
later insisted
\cite{Pauli} that ``the appearance of a
definite position of an electron during an observation is a {\it
creation} outside the laws of nature'' (my translation and italics).
Heisenberg had even claimed that ``the particle trajectory is created
by our act of observing it''. In accordance with Born's original ideas
(and also with von Neumann's orthodox interpretation), such spontaneous
``events'' are thus understood dynamically (in contrast to a mere
increase of information), while the process of observation or
measurement is not further dynamically analyzed.

\looseness-1
According to Heisenberg and the
early Niels Bohr, these individual {\it events} occurred in the atoms,
but this assumption had soon to be abandoned because of the existence
of larger quantum systems. Bohr later placed them into the
irreversible detection process. Others (such as London and Bauer
\cite{London} or Wigner
\cite{Wigner}) suggested that the ultimate events occur in the
observer, or that the ``Heisenberg cut'', where the probability
interpretation is applied in the observational chain between observer
and observed, is quite arbitrary. Ulfbeck and Aage Bohr \cite{UB}
recently wrote that ``the click  in the counter occurs out of the
blue, and without an event in the source itself as a precursor to the
click''. Note that there would then be no particle or other real
object any more that dynamically connects the source with the counter!
These authors add that simultaneously with the occurrence of the
click the wave function ``loses its meaning''. This is indeed the way
the wave function is often {\it used} -- though {\it not} whenever the
``particle'' is measured repeatedly (such as when giving rise to a
track in a bubble chamber). Even if it is absorbed when being measured
for the first time, the state thereafter is
described by a state vector that represents the
corresponding vacuum (which is evidently an {\it individually
meaningful} quantum state).

The picture that spontaneous {\it
events} are real, while the wave function merely describes their
deterministically evolving probabilities (as Born formulated
it) became general quantum folklore. It {\it could} represent an
objective description if these events were consistently described
by a fundamental stochastic process ``in nature'' -- for example in
terms of stochastic electron trajectories. A physical state at time
$t_0$ would then {\it incompletely} determine that at another time
$t_1$, say. The former could be said to contain ``incomplete
information'' about the latter in an objective dynamical sense (in
contrast to Heisenberg's concept of actual ``human knowledge'', or
information that is processed in a computer). This indeterminism would
be described by the spreading of a probability distribution,
representing the decay of ``objective information'', contained in an
initial state, about the later state. Unfortunately, this dynamical
interpretation fails. For example, it would be in conflict with
Wheeler's delayed choice experiments
\cite{frontiers}. Therefore, there have been attempts to reduce the
concept of trajectories to that of ``consistent
histories'' ({\it partially} defined trajectories)
\cite{Griffiths}. Roughly, these histories consist of successions
of discrete stochastic events that occur in situations being equivalent
to the aforementioned ``counters''. However, what circumstances let a
physical system qualify as a counter in this sense?

Can something that {\it affects} real events, or that keeps
solid bodies from collapsing, itself be unreal? In principle, this is
indeed a matter of definition. For example, electromagnetic fields were
originally regarded as abstract auxiliary concepts, merely useful to
calculate forces between the (``really existing'') charged elements of
matter. Bohm's theory demonstrates that electron trajectories
{\it can} be consistently {\it assumed} to exist, and even to be
deterministic under the guidance of a global wave function. Their
unpredictability would be due to unknown (and unknowable) initial
conditions. John Bell \cite{Bell} argued that the assumed
global wave function would then have to be regarded as real, too: it
``kicks'' the electron (while it is not being kicked back in this
theory). Evidently this wave function cannot {\it merely} represent a
statistical ensemble, although it dynamically {\it determines} an
ensemble of potential events (of which but one is supposed to {\it
become} real in each case -- note the presumed direction of time!).

In particular, any entanglement of the wave function is
{\it transformed} into statistical correlations
whenever (local) events {\it occur} without being observed. Even when
Schr\"odinger \cite{Schr} later called entanglement the greatest
mystery of quantum theory, he used the insufficient phrase
``probability relations in separated systems'' in the title to his
important paper. In the same year, Einstein, Podolski and Rosen
  concluded, also by
using entangled states, that quantum theory must be incomplete. The
importance of entanglement for the (real!) binding energy of the Helium
atom was well known by then, total angular momentum eigenstates were
known to require {\it superpositions} of products of subsystem states,
while von Neumann, in his book, had discussed the specific
entanglement that arises from quantum measurements. Nonetheless, none
of these great physicists was ready to dismiss the condition that
reality must be local (that is, defined in space and time). It is this
requirement that led Niels Bohr to abandon microscopic reality
completely (while he preserved this concept for the classical realm of
events).

\section{The Reality of Superpositions}
However, there seems to be more to the wave function than
its statistical and dynamical aspects. Dirac's general {\it
kinematical} concept of ``quantum states'' (described by his ket
vectors in Hilbert space) is based on the superposition
principle. It requires, for example, that the superposition of
spin-up and  spin-down defines a new {\it individual} physical
state, and  does not just lead to interference fringes in the
statistics of certain events. For every such spin superposition of
a neutron, say, there exists a certain orientation of a Stern-Gerlach
device, such that the path of the neutron can be {\it predicted with
certainty} (to use an argument from Einstein, Podolski and Rosen).
This spinor would not be correctly described  as a vector with two
unknown components.  Other spin components have to be
{\it created} (outside
the laws of nature according to Pauli) during measurements with another
orientation of the Stern-Gerlach magnet.

Superpositions of a neutron and a proton in the isotopic spin formalism
are formally analogous to spin, although the SU(2) symmetry
is dynamically broken in this case. As these superpositions do not
occur in nature as free nucleons (they may form quasi-particles within
nuclei), the validity of the superposition principle has been
restricted by postulating a ``charge superselection rule''. We can
now {\it explain} the non-occurrence of these and many other
conceivable but never observed superpositions by means of
envirenmental decoherence, while {\it neutral} particles, such as K
mesons and their antiparticles, or various kinds of neutrinos, {\it
can} be superposed to form new bosons or fermions, respectively, with
novel individual and observable properties.

Two-state superpositions {\it in space} can be formed from partial
waves which escape from two slits of a screen. Since these partial
waves can hardly be exactly re-focussed on  to the same point, we have
to rely on statistical interference experiments (using the probability
interpretation) to demonstrate the existence of this superposition of
single-particle wave functions. (The outcome of the required {\it
series} of events, such as sets of spots on a photographic plate, have
quantum mechanically to be described by tensor products of local states
describing such spots -- not by ensembles.) {\it General} one-particle
wave functions can themselves be understood as superpositions of all
possible ``particle'' positions (space points). They define ``real''
physical properties, such as energy, momentum, or angular momentum,
only as a nonlocal whole.

Superpositions of different particle numbers form another important
application of this basic principle, important for characterizing
``field'' states. If free fields are treated as continua of coupled
oscillators, boson numbers appear as the corresponding equidistant
excitation modes. Their coherent superpositions (which first appeared
in Schr\"odinger's attempt to describe corpuscles as wave packets) may
represent quasi-classical fields. Conversely, {\it quantum}
superpositions of classical fields define field functionals, that is,
wave functions over a configurations space of classical field
amplitudes.

These field functionals (generalized wave functions) were used
by Dyson \cite{Dyson} to derive path integrals and Feynman
diagrams for perturbation theory of QED. All particle lines in these
diagrams are no more than an intuitive short hand notation for plane
waves appearing in the integrals that are actually {\it used}. The
misinterpretation of the wave function as a probability distribution
for classical configurations (from which a subensemble can be ``picked
out'' by an increase of information) is often carried over to the
path integral. In particular, quantum cosmologists are
using the uncertainty relations to justify an {\it ensemble}
of initial states (an initial indeterminacy) for presumed trajectories
of the universe. Everett's relative state interpretation (based on the
assumption of a universal wave function) is then easily
misunderstood as a many-classical-worlds interpretation. However,
Heisenberg's uncertainty relations are defined with respect to {\it
classical} variables. They hold for {\it given} quantum states, and
do not require the latters' (initial, in this case) indeterminacy as
well.  Ensembles of quantum states would again have to be {\it
created} from an initial superposition (outside the laws or by means
of new laws), while {\it apparent} ensembles of cosmic fluctuations may
form by means of decoherence
\cite{K+Polarski}.

Superpositions of different states that are
generated from one asymmetric state by applying a symmetry group
(rotations, for example) are particularly useful. They
define irreducible representations (eigenstates of the corresponding
Casimir operators) as new individual physical states, which give rise
to various kinds of families of states or (real)
``particles''.

During the recent decades, more and more superpositions have
been confirmed to {\it exist} by clever experimentalists. We
have learned about SQUIDs, mesoscopic Schr\"odinger cats, Bose
condensates, and even superpositions of a macroscopic current running
in opposite directions (very different from two currents canceling
each other). Microscopic elements of quantum computers (which
simultaneously perform different calculations in one superposition)
have been successfully designed. All these superpositions may be (or
must be) observed as individual physical states. Hence, their
components  ``exist'' simultaneously. As
long as no unpredictable events have {\it occurred}, they
do {\it not} form ensembles of {\it possible} states which would
represent incomplete information about the true state.

A typical example for the {\it appearance} of
probabilistic quantum events is the decay of an unstable state by
means of tunneling through a potential barrier. The decay products
(which in quantum cosmology may even be whole universes) are here
assumed to enter existence, or to leave the potential well, at a
certain though unpredictable time. That this description is not
generally applicable has been demonstrated by experiments in cavities
\cite{decay}, where different decay times may interfere with one
another. Many narrow wave packets, approximately representing definite
decay times, would indeed have to be added coherently in order to form
unitarily evolving wave functions which may decay
exponentially (according to
a complex energy) in a large but restricted spacetime region.
This dynamical description by means of the Schr\"odinger equation
requires furthermore that exponential tails of an exact energy
eigenstate need time to form. So it excludes superluminal effects that
would result (though with very small probability) if exact eigenstates
were created in discontinuous quantum jumps
\cite{Hegerfeld}.

The conventional quantization rules, which are applied in
situations where a corresponding classical theory is known or
postulated, define the wave function $\psi(q)$ as a continuum of
coefficients in superpositions $\int\psi(q)|q \rangle$ of all classical
configurations $q$.\footnote{ Permutation symmetries and quantum
statistics demonstrate that the {\it correct} classical states to be
quantized are always spatial fields -- not the apparent
particles which appear as a consequence of decoherence. This has
recently been nicely illustrated by various experiments with Bose
condensates. The equidistant energy levels of free
field modes (oscillators) mimic particle numbers, and -- if robust
under decoherence -- give rise to the non-relativistic approximation by
means of ``$N$-particle wave functions''. This conclusion eliminates
any need for a ``second quantization'', while particle indices become
{\it redundant}, similar to gauge variables. {\it There are no
particles even before quantization} (while a fundamental unified
theory may not be based on any classical theory at all)!} With the
exception of single-particle states, this prodedure leads directly to
the infamous {\it nonlocal states}. Their nonlocality is very
different from a (classical) {\it extension} in space, which would
quantum mechanically be described by a {\it product} of local states.
Only {\it superpositions} of such products of subsystem states may be
nonlocal in the quantum mechanical sense.  In order to prevent reality
from being nonlocal, superpositions were therefore usually regarded as
states of information -- in contrast to the conclusion arrived at
above. Even hypothetical ``baby'' or ``bubble universes'' are defined
to be somewhere else {\it in space} -- in contrast to the ``many
worlds'' of quantum theory. However, Bell's inequality
  -- and
even more directly so its generalizations by Greenberger, Horne and
Zeilinger \cite{GHZ} or Hardy \cite{Hardy} -- have allowed
experimentalists to demonstrate by operational means that {\it reality}
is nonlocal. So why not simply accept the reality of the wave
function?

As mentioned above, there are {\it two}, apparently unrelated, aspects
which seem to support an interpretation of the wave function as a state
of ``information'': the classical {\it configuration} space, which
replaces normal space as the  ``stage of dynamics'' (and
thus leads to quantum nonlocality), and the probability interpretation.
Therefore, this picture and terminology appear quite appropriate for
practical purposes.  I am using it myself -- although preferentially
in quotation marks whenever questions of interpretation are discussed.

While the general superposition principle, from which nonlocality
is derived, requires nonlocal {\it states} (that is, a {\it
kinematical} nonlocality), most physicists seem to regard as
conceivable only a {\it dynamical} nonlocality (such as Einstein's
spooky action at a distance). The latter would even have to include
superluminal actions. In contrast, nonlocal entanglement must already
``exist'' before any crucially related local but spatially separated
events occur. For example, in proposed so-called quantum teleportation
experiments, a nonlocal state has to be carefully prepared initially
-- so nothing has to be {\it ported} any more. After this preparation,
the relevant state ``exists but is not there'' \cite{JZ}. Or in similar
words: the physical state is {\it ou topos} (at no place), although it
is {\it not utopic} according to quantum theory. If nonlocality is thus
readily described by the formalism (just taken
seriously), how about the probability interpretation?

\section{The R\^ole of Decoherence}
Most nonlocal superpositions discussed in the literature
describe controllable (or usable) entanglement. This is the reason why
they are being investigated. In an operationalist approach, this
usable part is often exclusively {\it defined} as entanglement, while
uncontrollable entanglement is regarded as  ``distortion'' or
``noise''. However, if the Schr\"odinger equation is assumed to be
universally valid, the wave function must contain far more
entanglement (or ``quantum correlations'') than can ever be used
\cite{Z1970}. In contrast to entanglement, uncontrollable noise, such
as phases fluctuating in time, would {\it not} destroy (or rather,
dislocalize) an individual superposition. It may at most wash out an
interference pattern in the {\it statistics} of many events (cf.\ts
\cite{Giulini}). Therefore, entanglement, which leads to decoherence
in bounded systems even for an {\it individual} global quantum state,
has to be strictly distinguished from phase averaging either in
ensembles of events, or resulting from a fluctuating Hamiltonian
(``dephasing'').

John von Neumann discussed the entanglement that arises when a quantum
system is measured by an appropriate device. It leads to the
consequence that the relative phases which characterize a
superposition are now neither in the object nor in the apparatus, but
only in their (shared) total state. These phases cannot affect
measurements performed at one or the other of these two subsystems any
more. The latter are then conveniently described by their reduced
density matrices, which can be formally {\it represented by ensembles}
of subsystem wave functions with certain formal probabilities. If the
interaction dynamics between system and apparatus is (according to von
Neumann) reasonably chosen, the resulting density matrix of the
apparatus by itself can be represented by an ensemble of slightly
overlapping wave packets which describe different pointer positions
with precisely Born's probabilities. Does it therefore explain the
required ensemble of measurement outcomes? That is, have the quantum
jumps into these new wave functions (the unpredictable events) already
occurred according to von Neumann's unitary interaction?

Clearly not. Von Neumann's model
interaction, which leads to this entanglement, can in principle be
reversed in order to reproduce a local superposition that depends on
the initial phase relation. For a microscopic pointer variable this
can be experimentally confirmed. For this reason,
d'Espagnat \cite{dE} distinguished conceptually between proper
mixtures (which describe ensembles) and improper mixtures (which are
defined to describe entanglement with an external system). This
difference is of utmost importance when the problem of measurement is
being discussed. The density matrix by itself is a formal tool that is
sufficient for all practical purposes which {\it presume} the
probability interpretation and neglect the possibility of local phase
revivals (recoherence). Measurements by {\it microscopic} pointers can
be regarded as ``virtual measurements'' (leading to ``virtual
decoherence'') -- in the sense of virtual particle
emission or virtual excitation. Similarly, scattering ``events''
cannot be treated probabilistically as far as phase relations,
described by the scattering matrix, remain relevant or {\it usable}.
Nothing can be assumed to have irreversibly happened in virtual
measurements (as it would be required for real events).

The concept of a reduced density matrix obtains its justification from
the fact that all potential measurements are local, that is,
described by local interactions with a local apparatus. Classically,
dynamical locality means that an object can {\it directly} affect the
state of another one only if it {\it is} at the same place. However,
we have seen that quantum states are at no place, in general. So what
does dynamical locality {\it mean} in quantum theory?

This locality (which, in particular, is a prerequisite for quantum
field theory) is based on an important structure that goes beyond the
mere Hilbert space structure of quantum theory. It requires (1) that
there is a Hilbert space {\it basis} consisting of local states
(usually a ``classical configuration space''), and (2) that the
Hamiltonian is a sum or spatial integral over corresponding local
operators. (The first condition may require the inclusion of gauge
degrees of freedom.) For example, the configuration space of a
fundamental quantum field theory is expected to consist of the
totality of certain classical field configurations on three- (or
more-) dimensional space, while its Hamiltonian is an integral over
products of these field operators and their derivatives. This form
warrants dynamical locality (in relativistic and nonrelativistic form)
in spite of the nonlocal kinematics of the generic quantum states.

So let us come back to the question why events and measurement results
appear actual rather than virtual. In order to answer it we
must first understand the difference between reversible and
irreversible (uncontrollable) entanglement. For this purpose we have
to take into account the realistic environment of a quantum system.
We may then convince ourselves by means of explicit estimates that a
macroscopic pointer cannot avoid becoming strongly entangled with its
environment through an uncontrollable avalanche of interactions, while
the quantum state of a microscopic variable remains almost
unaffected in most cases.

This situation has been
studied in much detail in the theory of decoher\-ence\footnote{The
concept of decoherence became known and popular through the ``causal''
chain Wigner-Wheeler-Zurek.} \cite{Giulini,
Zurek2001} (while many important applications still  remain to be
investigated -- for example in chemistry).  It turns out that all phase
relations between macroscopically different pointer positions become
irreversibly nonlocal in this way for all practical purposes and within
very short times -- similar to the statistical correlations that would
classically arise from Boltzmann's molecular collisions.
These chaotic correlations as well as the quantum
phases become inaccessible and irrelevant for the future evolution,
while they still exist according to the assumed deterministic
dynamics. If the wave function did ``lose meaning'', we would not
be able to {\it derive} decoherence from universal quantum dynamics.
 
The asymmetry in time of
this dissipation of correlations requires special initial conditions
for the state of the universe -- in quantum theory for its
wave function \cite{TD}.  However, in contrast to classical
statistical correlations, the arising entanglement
(``quantum correlation'') affects the individual state:  it represents
a formal ``plus'' rather than an ``or'' that would characterize an
ensemble.

Two conclusions have to be drawn at this point. Decoherence occurs
according to the reversable dynamical laws (the Schr\"odinger
equation) by means of an in practice irreversible process, and
precisely where events {\it seem} to occur, but even this success does
{\it not} lead to an ensemble representing incomplete information. The
improper mixture does {\it not} become a proper one. We can neither
unambiguously choose a specific ensemble to ``represent'' the reduced
density matrix, nor even the  subsystem to which the latter belongs.

 From a {\it fundamental} point of view it would, therefore, be
misleading in a twofold way to regard the entangled wave function as
representing ``quantum information''. This terminology suggests
incorrectly (1) the presence of a (local) reality that is
incompletely described or predicted by the wave function, and (2) the
irrelevance of environmental decoherence  for the measurement process
(even though it has been experimentally confirmed
\cite{haroche}).

A further dynamical consequence of decoherence is essential for the
pragmatic characterization of the observed classical physical world.
Consider a two-state system with states $|L>$ and $|R>$ which are
``continually measured'' by their environment, and assume that they
have exactly the same diagonal elements in a density matrix. Then this
density matrix would be diagonal in any basis after complete
decoherence. While a very small deviation from this degeneracy would
soon resolve this deadlock, an exact equality {\it could} arise from
an exactly symmetric initial state,
$|\pm> = (|R>
\pm |L>)/\sqrt{2}$. However, if we then
measured
$|R>$, say, a second measurement would confirm this result, while a
measurement of
$|+>$ (if possible) would give $|+>$ or
$|->$ with equal probabilities when repeated after a short decoherence
time. It is the ``robustness'' of a certain basis  under decoherence
(a ``predictability sieve'' in Zurek's language)
that
gives rise to its classical appearance. In the case of a measurement
apparatus it is then called a ``pointer basis''.

\looseness-1
This problem of degenerate probabilities remains present also for
quasi-degenerate {\it continua} of states. For sufficiently massive
particles (or macroscopic pointer variables), narrow wave packets may
be robust even though they do not form an orthogonal basis that
diagonalizes the density matrix. Their shape and size may change under
decoherence in spite of their robustness. Collective
variables (such as the amplitude of a surface vibration) are
adiabatically ``felt'' (or ``measured'') by the individual particles.
For microscopic systems this would represent a mere dressing of the
collective mode. (My original work on decoherence was indeed
influenced by John Wheeler's work on collective nuclear vibrations by
means of {\it generator coordinates}
\cite{GW}.) However, decoherence must be irreversible. Even
the germs of all cosmic inhomogeneities were irreversibly ``created''
by the power of decoherence in breaking the initial
homogeneity during early inflation
\cite{K+Polarski}. In other cases, such as a gas under normal
conditions, localized molecules {\it seem} to exist as a consequence
of decoherence into narrow wave packets, while the latters' lacking
robustness prevents the formation of extended individual trajectories.

Nonetheless, something is still missing in the theory in order to
arrive at {\it definite} events or outcomes of measurements, since the
{\it global} superposition must still exist according to the
Schr\"odinger equation. The most conventional way out of this dilemma
would be to postulate an appropriate collapse of the wave function as
a fundamental modification of unitary dynamics. Several models have
been proposed in the literature
(cf.\ts \cite{GRW}). They (quite unnecessarily) attempt to mimic
precisely the observed environmental decoherence. However, since
superpositions have now been confirmed far in the macroscopic realm, a
Heisenberg cut for the application of the collapse may be placed {\it
anywhere} between the counter (where decoherence first occurs in the
observational chain) and the observer. The definition of subsystems in
the intervening medium is entirely arbitrary, while the
diagonalization of their reduced density matrices (the choice of their
``pointer bases'') may be convenient, but is actually irrelevant for
this purpose. An individual observer may even ``solipsistically''
assume the border line to exist in the observational chain after
another human observer (who is usually referred to as ``Wigner's
friend'', since Eugene Wigner first discussed this situation in the
r\^ole of the final observer).

It would in fact not help very much to postulate a collapse to occur
somewhere in the counter. The physical systems which carry the
information from the counter to the observer, including his sensorium
and even his brain, must all be described by quantum mechanics.
Quantum theory applies everywhere, even where
decoherence allows it to be approximately replaced by stochastic
dynamics in terms of quasi-classical
concepts (``consistent histories''). In an important paper, Max
Tegmark \cite{tegmark} recently estimated that neuronal networks and
even smaller subsystems of the brain are strongly affected by
decoherence. While this result does allow (or even requires)
probabilistic quantum effects, it excludes extended controllable
superpositions in the brain, which might represent some kind of
quantum computing. However, postulating a probability interpretation
at this point would eliminate the need for postulating it anywhere
else in the observational chain. It is the (local) {\it
classical} world that seems to be an illusion!

Nobody knows as yet where precisely (and in fact whether) consciousness
may be located as the ``ultimate observer system''. Without
any novel empirical evidence there is no way to decide where a
collapse really occurs, or whether we have indeed to assume a
superposition of many classical worlds -- including ``many minds''
\cite{ZConsc} for each observer -- in accordance with a {\it universal}
Schr\"odinger equation. It is sufficient for all practical purposes to
know that, due to the irreversibility of decoherence, these different
minds are dynamically autonomous (independent of each other) after an
observation has been completed. Therefore, Tegmark's
quasi-digitalization of the neuronal system (similar to the $|R,L>$
system discussed above) may even allow us to define this subjective
Everett branching by means of the diagonal form of the observer
subsystem density matrix.

A genuine collapse (in the counter, for example) would produce an
unpredictable result (described by one {\it component} of the
wave function that existed prior to the collapse). The state of
ignorance after a collapse with unobserved outcome is, therefore,
described by the {\it ensemble} of all these components with
corresponding probabilities. In order to reduce this ensemble by an
increase of information, the observer has to interact with the
detector in a quasi-classical process of observation. In the many-minds
interpretation, in contrast, there is an objective process of
decoherence that does {\it not} produce an ensemble. (The reduced
density matrix resulting from decoherence can be treated for all
practical purposes {\it as though} it represented one. This explains
the {\it apparently observed collapse} of the wave function.) Even the
superposition of the resulting many minds describes {\it one} quantum
state of the universe. Only from a subjective (though objectivizable
by entanglement) point of view would there be a transition into {\it
one} of these many ``minds'' (without any intermediary ensemble in
this case). This interpretation is reminiscent of {\it Anaxagoras'
doctrine}, proposed to separate Anaximander's  {\it apeiron} (a state
of complete symmetry): ``The things that are in a single world are not
parted from one another, not cut away with an axe, neither the warm
from the cold nor the cold from the warm. When Mind began to set
things in motion, separation took place from each thing that was being
moved, and all that Mind moved was separated.'' (Quoted from
\cite{Jammer}, p.\ts 482). Although, according to this quantum
description, the r\^ole of ``Mind'' remains that of a passive (though
essential) epi-phenomenon (that can never be {\it explained} in terms
of physical conepts), we will see in the next section how Anaxagoras'
``doctrine'' would even apply to the concepts of motion and time
themselves.
 
\looseness-1
In this specific sense one might introduce the {\it terminology}
(though {\it not} as an explanation) that the global wave function
represents ``quantum information''. While decoherence transforms the
formal ``plus'' of a superposition into an effective ``and'' (an {\it
apparent} ensemble of new wave functions), this ``and'' becomes an
``or'' only with respect to a subjective observer. An {\it additional}
assumption has still to be made in order to justify Born's
probabilities (which are meaningful to an individual mind in the form
of frequencies in {\it series} of measurements): one has to assume
that ``our'' (quantum correlated) minds are located in a component of
the universal wave function with non-negligible norm
\cite{graham}. (Note that this is a {\it probable} assumption only
after it has been made.) It is even conceivable that observers may not
have been able to evolve at all in other branches, where
Born's rules would not hold
\cite{Saunders}.

\section{The Wheeler-DeWitt Wave Function}
The essential lesson of decoherence is that the whole universe
must be strongly entangled. This is an unavoidable
consequence of quantum dynamics under realistic assumptions. In
principle, we would have to know the whole wave function of the
universe in order to make local predictions. Fortunately, there are
useful local approximations, and most things may be neglected in most
applications that are relevant for us local observers. (Very few
systems, such as the hydrogen atom, are sufficiently closed and simple
to allow precision tests of the theory itself.)

For example,  gravity seems to be negligible in most situations, but
Einstein's metric tensor defines space and time -- concepts which are
always relevant. Erich Joos \cite{Joos} first argued that the quantized
metric field is strongly decohered by matter, and may therefore usually
be treated classically. However, some aspects of quantum gravity are
essential from a fundamental and cosmological point of view.

General relativity (or any unified theory containing it) is
invariant under reparametrization of the (physically meaningless) time
coordinate
$t$ that is used to describe the dynamics of the metric tensor. This
invariance requires trajectories (in the corresponding configuration
space) for which the Hamiltonian vanishes. This {\it Hamiltonian
constraint},
$H=0$, can thus classically be understood as a conserved initial
condition (a conserved ``law of the instant'') for the time-dependent
states. Upon quantization it assumes the form of the {\it
Wheeler-DeWitt equation} (WDWE),
$$
H\Psi = 0 \quad ,
$$
as the ultimate
Schr\"odinger equation \cite{DWW}. This wave function $\Psi$ depends on
all variables of the universe (matter and geometry, or any unified
fields instead). Since now $\partial_t \Psi = 0 $, the static
constraint is all that remains of dynamics. While the classical law of
the instant is compatible with time dependent states (trajectories),
time is entirely lost on a fundamental level according to the WDWE.
For a wave function that describes reality, this result cannot be
regarded as just formal. ``Time is not primordial!''
\cite{frontiers}

Dynamical aspects are still present, however, since the WDW wave
function
$\Psi$ describes entanglement between all variables of the universe,
including those representing appropriate clocks. Time dependence is
thus replaced by quantum correlations \cite{PW}. Among these variables
is the spatial metric (``three-geometry''), which defines time as a
{\it many-fingered controller of motion} for matter
\cite{BSW} (another deep conceptual insight of John Wheeler), just as
Newton's time controls motion  in an absolute sense.

The general solution of this WDWE requires cosmic boundary
conditions in its configuration space. They may not appear very
relevant for ``us'', since $\Psi$ describes the superposition of
``many worlds''. Surprisingly, for Friedmann type universes, this
static equation is of hyperbolic type after gauge degrees of freedom
have been removed: the boundary value problem becomes an initial value
problem with respect to the cosmic expansion parameter $a$ \cite{Z86}.
For appropriate boundary conditions at $a=0$, this allows one to deduce
a cosmic arrow of time (identical with that of cosmic expansion)
\cite{TD}. However, in the absence of external time $t$, there is
neither any justification for interpreting the wave function of the
whole universe in a classically forbidden region as describing a
tunneling {\it process} (or the probability for an {\it event}), nor
to distinguish between its expansion and contraction according to the
phase of $e^{\pm ia}$ (see
\cite{Vilenkin}, for example). In contrast, an
$\alpha$-particle is found with an outgoing wave after a metastable
nuclear state has been {\it prepared} (in external time). Similar
arguments as for a tunneling process apply to a classical ``slow
roll'' of the universe along a descending potential well
\cite{ST}.

Since the Wheeler-DeWitt wave function represents a superposition of
all three-geometries (entangled with matter), it does not
describe quasi-classical histories (defined as
one-dimensional successions of states, or instants). Kiefer was able
to show \cite{Kiefer} that such histories (which define spacetimes)
can be approximately recovered by means of decoherence along WKB
trajectories that arise according to a Born-Oppen\-heimer
approximation with respect to the Planck mass. This leads to an
effective time-dependent Schr\"odinger equation along each WKB
trajectory in superspace (Wheeler's term for the configuration  space
of three-geo\-metries). Complex branch wave functions emerge thereby
from the real WDW wave function by an intrinsic breaking of the
symmetry  of the WDWE under complex conjugation (cf.\ts Sect.\ts 9.6
of
\cite{Giulini}). Each WKB trajectory then describes a whole
(further branching) Everett universe for matter.

Claus Kiefer and I have been discussing the problem of timelessness
with Julian Barbour (who wrote a popular book \cite{Barbour} about it)
since the mid-eighties. Although we agree with him that time can only
have emerged as an approximate concept from a fundamental timeless
quantum world that is described by the Wheeler-DeWitt equation, our
initial approach and even our present understandings differ. While
Barbour regards a {\it classical} general-relativistic world as
time-less, Kiefer and I prefer the interpretation that timelessness is
a specific quantum aspect (since there are not even parametrizable
trajectories in quantum theory). In classical general relativity, only
{\it absolute} time (a preferred time parameter) is missing, while the
concept of one-dimensional successions of states remains valid.

In particular, Barbour regards the {\it classical} configuration space
(in contradistinction to the corresponding momentum space or to phase
space) as a space of global actualities or ``Nows''. {\it Presuming}
that time does not exist (on the basis that there is no absolute
time), he then {\it concludes} that trajectories (of which but one
would be real in conventional classical description) must be replaced
by the multi-dimensional continuum of {\it all} potential Nows (that
he calls ``Platonia''). He assumes this continuum to be
``dynamically'' controlled by the WDWE. After furthermore presuming a
probability interpretation of the WDW wave function for his global
Nows (in what may be regarded as a {\it Bohm theory without
trajectories}), he is able to show along the lines of Mott's
theory of
$\alpha$-particle tracks (and by using Kiefer's results) that classical
configurations which are considerably ``off-track'' (and thus without
memory of an {\it apparent} history) are extremely improbable. Thus
come memories without a history.

One might say
that according to this interpretation the Wheeler-DeWitt wave function
is a {\it multidimensional generalization of one-dimensional
time} \cite{TD}. Julian Barbour does not agree with this terminology,
since he insists on the complete absence of time (although this may be
a matter of words). I do not like this picture too much for other
reasons, since I feel that global Nows are not required, and that the
Hamiltonian symmetry between configuration space and momentum space is
only dynamically -- not conceptually -- broken (by dynamical
locality). Nonetheless, this is a neat and novel idea that I feel
is worth being mentioned on this occasion.

\section{That ITsy BITsy Wave Function}
Reality became a problem in quantum theory when physicists desperately
wanted to understand whether the electron or the photon ``really'' is
a particle or a wave (in space). This quest aimed at no more than a
{\it conceptually consistent description} that may have to be
guessed, but would then be confirmed by all experiments. It was
dismissed in the Copenhagen interpretation according to the program  of
complementarity (which has therefore also been called a
``non-concept'').  I have here neither argued for particles nor for
spatial waves, but instead for Everett's (nonlocal) wave function(al).
It may serve as a consistent and universal kinematical concept, and in
this sense as a description of reality  (once supported also by John
Wheeler
\cite{WhRevModPhys}).  The price may appear high: a vast multitude of
separately observed (and thus, whith one exception, unobservable to us)
quasi-classical universes in one huge superposition. However, similar
to Everett I have here no more than extrapolated those concepts which
are successfully {\it used} by quantum physicists. If this
extrapolation is valid, the price would turn into an enormous dividend
of grown knowledge about an otherwise hidden reality.

The concept of reality has alternatively been based on operationalism.
Its elements are then defined by means of operations (performed by what
Wheeler called ``observer-participators'' \cite{frontiers}), while
these operations are themselves described in non-technical
``every-day'' terms {\it in space and time}. In classical physics,
this approach led successfully to physical concepts which proved
consistently applicable. An example is the electric field, which was
defined by means of the force on (real or hypothetical) test
particles. The required operational means (apparata) could afterwards
be self-consistently described by using these new concepts
themselves (partial reductionism). This approach fails in quantum
theory, since, for example, quantum fields would be strongly affected
(decohered) by test particles.

The investigation of quantum objects thus required various, mutually
incompatible, operational means. This led to
incompatible (or ``complementary'') concepts, seemingly in
conflict with a microscopic reality.
Niels Bohr's ingenuity allowed him to recognize this situation very
early. Unfortunately, his enormous influence (together with the dogma
that the concept of reality must be confined to objects in space and
time) seems to have prevented his contemporaries to {\it explain} it in
terms of a more general (nonlocal) concept that is successfully {\it
used} but  not directly accessible by means of operations: the
universal wave function. In terms of this hypothetical reality  we may
now understand  why certain (``classical'') properties are robust even
when being observed, while microscopic objects may interact with
mutually exclusive quasi-classical devices under the control of clever
experimentalists. However, this does {\it not} mean that these quantum
objects have to be {\it fundamentally} described by varying and
incompatible concepts (waves or particles, for example).

If ``it'' (reality) is understood in the operationalist sense, while
the wave function is regarded as ``bit'' (incomplete knowledge
about the outcome of potential operations), then
one or the other kind of {\it it} may indeed emerge {\it from bit}
-- depending on the ``very conditions'' of the operational situation.
I expect that this will remain the pragmatic language  for
phyicists to describe their experiments for some time to come. However,
if {\it it} is required to be described in terms of  not necessarily
operationally accessible but instead universally valid concepts, then
the wave function remains as the only available candidate for {\it
it}. In this case, {\it bit} (information as a dynamical functional
form, as usual) may emerge {\it from it}, provided an appropriate
(though as yet incompletely defined) version of a psycho-physical
parallelism is postulated in terms of this nonlocal {\it it}.  If
quantum theory appears as a ``smokey dragon'' \cite{frontiers}, the
dragon itself may now be recognized as a universal wave function,
partially veiled to us local beings by the ``smoke'' of its inherent
entanglement.

However you turn it: {\it In the Beginning Was the Wave Function}.
We may have to declare victory of the Schr\"odinger over the
Heisenberg picture.

\end{document}